\documentclass[twocolumn]{aastex63}

\usepackage{float}

\usepackage{enumitem} 
\usepackage{lipsum}
\usepackage{multirow}  
\usepackage{lineno}
\usepackage{tipa}
\usepackage{amsmath}
\usepackage{tabularx} 
\usepackage{array}

\shortauthors{}

\graphicspath{{./}{figures/}}

\received{April 1st, 2025}
\revised{April 1nd, 2025}
\accepted{April 1th, 2025}
\begin{document}

\title{The Rizzeta Stone: Adopting Gen-$\alpha$ Colloquial Language to Improve Scientific Paper Rizz and Aura from a Skibidi Perspective}

\author[0000-0002-8195-0563]{A. E. Blackwell}
\affiliation{University of Michigan, Ann Arbor, MI 48104, USA}

\author[0000-0003-1463-8702]{D.~L.~Moutard}
\affiliation{University of Michigan, Ann Arbor, MI 48104, USA}

\author[0000-0002-8195-0563]{J. A. Miller}
\affiliation{University of Texas A\&M, College Station, TX}

\begin{abstract}
The field of astronomy evolves rapidly, and it is essential to keep up with these changes in order to effectively communicate with the broader community. However, communication itself also changes as new words, phrases, and slang terms enter the common vernacular. This is especially true for the current youngest generations, who are capable of efficiently communicating via the Internet. In order to maintain effective communication, we explore the possibility of expanding the language used in scientific communication to include recently coined slang. This attempt at outreach, while potentially very difficult, could provide a means to expand the field and capture the attention of early-career scientists, improving retention within the field. However, our results indicate that, while possible, this method of communication is, like, probably not really worth it, no cap.   
\end{abstract}


\section{Introduction}

Astronomy and Astrophysics is an ever changing field with new ideas and community members introduced every day.
Subsequently, the field is subject to new advancements and discoveries that we all must adapt to. 
Some such examples include the discovery of dark matter \citep{zwicky1933}, the cosmic microwave background \citep[CMB,][]{penzias1965}, and the first confirmed exoplanet \citep{Wols1992}. 
All of these findings had remarkable implications. Dark matter is included in many large-scale computer simulations and has become its own subfield. 
We see evidence of the CMB in radio spectra of galaxy groups and clusters \citep[e.g.,][]{Mroczkowski2012} and efforts to create all-sky CMB maps are still underway \citep{Planck2016, chandran2023, pratt2024}. 
Finally, the first confirmed exoplanet was discovered by \cite{Wols1992} in 1992. Since then, the number of confirmed exoplanets has skyrocketed to $\sim$7500\footnote{https://exoplanet.eu/catalog/} and is now one of the fastest growing subfields in astronomy. 

One much less discussed change is the advent of new rhetoric, colloquially referred to as `slang'. 
Communication, specifically language, is the cornerstone of human interaction and provides a means of conveying thoughts, emotions, ideas, and new science. 
It is constantly evolving to accommodate social changes, technological advancements, and the blending of cultures. 
Although we see this evolution exceedingly in the use of slang, the incorporation of new terms in an academic setting is greatly lacking. 
This is in stark contrast to the field's incorporation of new scientific ideas. 
We suggest the incorporation of slang into academic publications, where its use may be seen more broadly and therefore be more inclusive to new generations joining the field. 

\begin{figure*}
    \centering
    \includegraphics[width=1\linewidth]{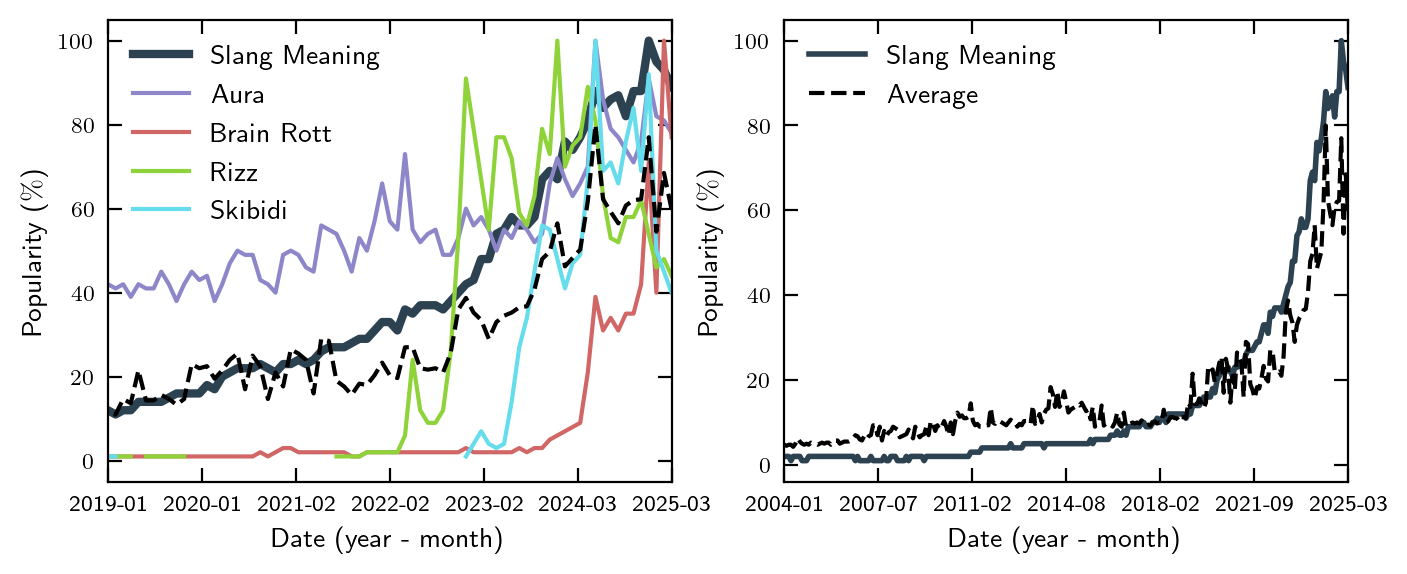}
    \caption{Google search history popularity as a function of time. In the right panel we show the trends of four recently popular slang words, aura (purple line), brain rot (red line), rizz (green line), and skibidi (blue line), with the dashed line representing the average of these 4 trends. We also show the Google search popularity of `slang meaning' (thick grey line). In the left panel we show a larger date range of the average trend compared to `slang meaning'. The average trend and that of `slang meaning' agree very well indicating a general interest in learning the new slang words as they are developed and popularized.}
    \label{fig:google_searches}
\end{figure*}

The demand for the use of slang has arisen from the astronomical increase in the rate of slang word development.
Figure \ref{fig:google_searches} shows the Google search history of recent slang words such as `aura', `brain rot', `rizz', and `skibidi'. 
All trends resemble an exponential function after the words invention and subsequent circulation. 
The advent of new slang words and their increasing popularity coincides with the general public's attempt to understand, and potentially use, these words.
This can be seen in the Google search history for 'slang meaning', which closely follows the trend of the Google search history for specific new slang words, and their average Google search rate (right and left panels of Figure \ref{fig:google_searches}, respectively).

The recent exceptional increase in slang words can be credited to social media. 
Platforms such as TikTok, YouTube, X (formerly Twitter), and Instagram enable the instant sharing of new terms and expressions globally. 
This results in the more homogenized, yet diverse, linguistic landscape we see now. 
However, it is crucial to note that while these platforms enable widespread diffusion, they also allow for the emergence of hyper-localized slang that may resonate only within niche communities. 
Much of the slang discussed in this paper originates from Generations Z and $\alpha$, birth years 1995$\sim$2011, and $\gtrsim$2012 respectively\footnote{https://libguides.usc.edu/busdem/age}. 

A large disconnect can occur between generations if older people do not frequently associate with members of the Generation $\alpha$ community. 
This unassociated or disconnected group is frequently referred to as the ``Boomers”, birth years 1946-1964.

When incorporating slang into your academic publications, it is important to consider the context within which a slang term is generated. 
Slang often emerges from specific cultural, social, or geographical environments, and its use is deeply embedded in the experiences and values of these communities. 
The context influences not only how the term is understood but also how it evolves over time. 
Misinterpretation can occur when the subtleties that gave rise to the term are overlooked. 
For example, ``skibidi Ohio” is frequently used to convey a large dislike, disinterest, or cringe feeling toward a subject. 
Someone who associates the American state of Ohio with positive feelings may misinterpret the speaker as having positive feelings about the subject. 
Or, they may take the term offensively. 
Therefore, we do not recommend incorporating ``skibidi Ohio” for fear of ostracizing colleagues at The Ohio State University.

In this work, we present ideas and suggestions for including slang in your future Astronomy and Astrophysics publications to enhance engagement with broader audiences. 
We provide a non-comprehensive overview on some of the most popular slang words of the past several years, and provide recommendations and examples on how to incorporate them into future academic works so that you too can stay hip.

The structure of our paper is as follows. 
In Section \ref{sec:methods} we outline the sources of our slang words.
We provide a brief etymology and table of slang words and their definitions in Section \ref{sec:slang}. 
Examples of `rizzified' sentences from classic papers from Astronomy and Astrophysics are provided in Section \ref{sec:apply}.
In Section \ref{sec:conclsion} we re-emphasize our recommendations on including slang in your future applications, and provide a sentiment of confidence so you may cook in future papers.

\section{Data Collection} 
\label{sec:methods}

Our data originates from a multi-year multi-messenger observation campaign of frequently utilized social media websites. These websites include, but are not limited to: Facebook, X (formerly Twitter), Instagram, Reddit, Tumblr, Tiktok, YouTube, and Bluesky. Observations were taken over the course of several years to better characterize the origination and evolution of slang terms across cosmic time. These observations began roughly 15~years ago and have been maintained with a frequent and stable observational cadence since. 

To corroborate our findings, we consulted two professionals in the education field that interact with youth who also frequently use these social media sites. They have both confirmed that the words utilized are current as of the publication of this manuscript. These professionals are located across the continental United States, so our data is only relevant for predominantly English speaking countries. However, this in itself speaks to how quickly slang terms can see widespread adoption through the use of social media. 

We caution that, while the spread of slang words is faster than ever through the use of the internet, this speed is not instantaneous. The terminology we present in this manuscript will have a significant bias towards the north-eastern regions of the United States. Other areas of the United States and the rest of the world will inevitably have their own slang. We attempt only to cover the most widespread and frequently used words relevant to serious scientific analyses.

\section{Slang Origins and Development} 
\label{sec:slang}

Here we present a summarized, alphabetized, noninclusive list of adopted slang words and their definitions, categorized by their connotative ``goodness''. We also discuss some of the cultural origins of modern slang words.
\begin{table*}
\center
\begin{tabular}{p{3cm} | p{4cm}| p{4cm} | p{4cm} } 
Term & Definition & Sample Sentence & Usage Notes \\
\hline
\hline
Aura \textit{n.} & A likeable/imposing demeanor, often determined just by appearance & \textit{``Bro has such \textbf{aura}, you see everybody laughing around him?"} & ``Aura Farming" is a common usage, attempting to look cool frequently without actually doing anything \\
GOAT \textit{n.} & Acronym for ``Greatest of all Time" & \textit{``Michael Jordan is the \textbf{GOAT} of basketball."} & \\
Gucci \textit{adj.} & Good; in good standing & \textit{``Don't worry about paying me back man, we're \textbf{Gucci.}"} & Originates from the high-end clothing retailer.\\
Gyatt \textit{n.} & An attractive buttocks & \textit{``Did you see that \textbf{gyatt}? [REDACTED]"} & Not suitable for serious academic use.\\
King/Queen \textit{n.} & Gendered terms signifying high status & ``\textbf{king}", ``go off \textbf{queen}" & Also used as an adjective to describe activities befitting such a person. \\
Rizz \textit{n.} & Derived from \textit{charisma}, an ability to sway others, particularly in the context of courtship and dating & \textit{``She's got unimaginable \textbf{rizz}; she came out of the bar with ten phone numbers."} & Also applicable as a verb, e.g. rizz/ing to describe the act of swaying others\\
Oomf \textit{n.} & Acronym for ``One of my followers"; a friend met through social media & \textit{``My \textbf{oomf} posted this extremely interesting paper on black holes!"} & \\
Skibidi \textit{adj.} & General term meaning ``Good" or something positive, occasionally used as a filler word. & \textit{``My boyfriend bought me my favorite flowers, he is so \textbf{skibidi}!"} & From YouTube series Skibidi Toilet, the word has since decoupled from the series.\\
\end{tabular}
\caption{Non-exhaustive list of positively-oriented slang terms. From left to right, the columns are the Term, definition of the term, an example sentence with the word bolded, and finally any additional notes relevant to its usage.}
\label{table:Slang_Positive}
\end{table*}
\begin{table*}
\center
\begin{tabular}{p{3cm} | p{4cm}| p{4cm} | p{4cm} } 

Term & Definition & Sample Sentence & Usage Notes \\
\hline
\hline
Chat \textit{n.} & A group of omniscient observers & \textit{``Can you believe this \textbf{chat}?"} & Originated from video game streamers referring to their audience, often used to imitate them. \\
Cook \textit{v.} & To continue to ac towards a strictly positive or negative outcome & \textit{``He's almost got it, let him \textbf{cook!}"} & While ``cook" is typically used to describe action toward a positive outcome, to be ``cooked" \textit{adj.} implies the opposite, i.e. to be ``defeated" or ``finished".\\
High-Key \textit{adv.} & Without subtlety & \textit{``I'm \textbf{high-key} excited for the concert tonight!"} & While its antonym ``low-key" means ``subtlety", high-key is typically used as an intensifier. \\
Lore \textit{n.} & The backstory or background information of an entity & \textit{``Stephanie has been through more than you would think, her \textbf{lore} is insane."} & \\
Mad \textit{adj.} & A lot of; a great deal of something & \textit{``Lizzy has \textbf{mad} skills when it comes to understanding quantum computers!}" & \\
Mid \textit{adj.} &Unremarkable or average & \textit{"Steven claims to be great at video games, but in actuality he is pretty \textbf{mid.}"} & While by definition not an insult, typically used in a derogatory manner \\
On-God \textit{adv.} & Truthful, particularly when a claim is faced with doubt or appears far-fetched & \textit{``\textbf{On-God}, I caught an alligator two feet long in my backyard last week!"} & \\
Out of Pocket \textit{adj.}  & Random or unexpected, but usually in a lighthearted manner & \textit{``Sarah's comment on bird mating habits was very \textbf{out of pocket} given we were discussing neutron stars, but it was pretty funny."} & \\
\end{tabular}
\caption{Non-exhaustive list of neutral-oriented slang terms.}
\label{table:Slang_Neutral}
\end{table*}
\begin{table*}
\center
\begin{tabular}{p{3cm} | p{4cm}| p{4cm} | p{4cm} } 
Term & Definition & Sample Sentence & Usage Notes \\
\hline
\hline
Brainrot \textit{n.} & The resultant state of an overuse of social media, typically presenting as having a short attention span and a vocabulary comprised mainly of internet slang & \textit{``Carol needs to lay off TikTok, her \textbf{brainrot} has gotten so bad I can hardly understand her."} & While often used derisively, it can also describe the genre of humor that utilizes `brainrot' slang. \\
Cap \textit{n.} & Lie; an un-truth & \textit{``George said they're coming to the party, but that's \textbf{cap}."} & Often phrased ``No Cap" to indicate honesty. \\
Crash Out \textit{v.} & To go crazy or berserk, typically in response to negative stimulus & \textit{``If I fail this next geometry test I swear I am going to \textbf{crash out}."} & \\
Glaze \textit{v.} & To hype up or compliment to the point of being a suck-up. & \textit{``Charles always \textbf{glazes} the boss's work in meetings, obviously he is eyeing that promotion."} & \\
Ohio \textit{adj.} & Weird; Strange & \textit{``I feel sick to my stomach, that lunch was definately made in \textbf{Ohio}."} &  \\
Opp \textit{n.} & A person opposed to someone; an enemy & \textit{``Frankie's \textbf{opps} are praying on his downfall."} & Typically used to refer to a general force of opposition, not a single person.\\
Sus \textit{adj.} & Untrustworthy or suspicious & \textit{``That guy in the trenchcoat behind the hardware store is real \textbf{sus}."} &  \\
Unc \textit{n.} & Short for ``uncle", refers to an older person than the speaker, typically with a negative connotation implied. & \textit{``Look at \textbf{unc}! Despite his knee surgery, he does still have some skill left in him."} & \\
\end{tabular}
\caption{Non-exhaustive list of negatively-oriented slang terms.}
\label{table:Slang_Negatve}
\end{table*}

\subsection{Etymology}

When considering the introduction of new terminology into communications for a specific field, it is important to consider their origins. While up to this point, we have considered broad systems such as ``the Internet" or ``Social Media" as the origins of these words or phrases, we must also remember that these systems begin and end with {\it people}. Specifically, we would like to highlight that the bulk of the slang introduced in this paper stems largely from under-represented groups, such as black Americans and young women. For example, terms such as ``cap" and ``gyatt" (among others) originate from African American Vernacular English (AAVE) \citep{sole24a,sole24b}, but have been disseminated to the broader public via social media and art forms pioneered by black Americans. Even the now-common word ``cool" meaning ``popular, good, worthwhile" stems from the African American led jazz movements of the 1940's \citep{cincotta18}. 

Young women are also frequently the progenitors of popular slang. For example, one can see this by looking back to the ``Valley Girl" style of speech popular among affluent young women on the West Coast of the United States in the 1980's. While much of the slang from this particular sociolect has fallen out of use, one in particular remains common to the time of writing -- using ``like" as a discourse marker or filler word \citep{keelty21}. For a more modern example, we can see the phrase Hawk Tuah\footnote{We have intentionally chosen not to define this phrase in this study to maintain accessibility to all ages. Defining this is left as an exercise to the reader.} which began as a young woman making a bawdy joke in a now-viral interview. For more discussion, see articles such as ''Teenage Girls Have Led Language Innovation for Centuries"\footnote{https://www.smithsonianmag.com/smart-news/teenage-girls-have-been-revolutionizing-language-16th-century-180956216/} and sources within.

Many of these words and phrases are considered ``incorrect" speech, or are otherwise looked down upon by large swathes of English speakers at the time of their inception. However, after some time, a shift happens and the speech becomes standard accepted slang, and perhaps eventually regular colloquial speech. It is in this light that we must recognize that a prescriptivist approach to defining what language is appropriate for particular settings is often rooted in a status quo that ignores or denigrates marginalized communities. This fact should be considered if an author wishes to utilize slang, if they wish to maintain respect for colleagues from communities who have been discouraged in the past from using the same language.
\newpage

\subsection{Development}

\begin{figure}
    \centering
    \includegraphics[width=1.0\linewidth]{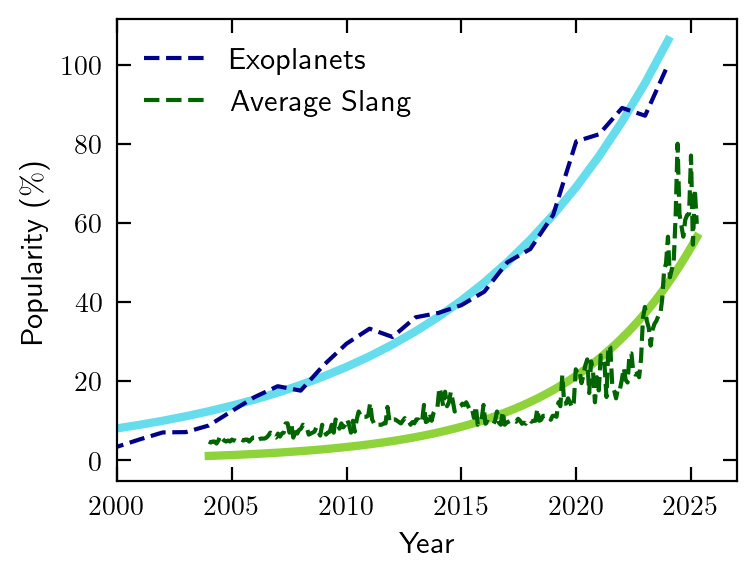}
    \caption{Popularity trends of the slang word averaged search (green dashed), and `exoplanet' in abstracts of papers submitted to {\sc ar$\chi$iv} (dashed blue line). Each trend is fit with an exponential function (solid line and respective color). The trend for average slang is steeper with an exponential growth rate of $0.18 \pm 0.007$, compared to the increase of `exoplanet' use with a growth rate of $0.11 \pm 0.003$.}
    \label{fig:trend_fits}
\end{figure}

Once these words do enter the public awareness, their growth can be rapid, often leaving behind the previous connotations of ``incorrect speech'' described above. We find in general that the increase in slang terminology roughly follows an exponential growth of the simple form 
\begin{equation}
    f(x) \propto e^{gt}
\end{equation}

where $g$ is the growth factor, with higher values indicating faster growth. As shown in Figure \ref{fig:trend_fits}, the average growth rate has a value $g \sim 0.18 \pm 0.007$. Compare this to the growth rate of the term ``exoplanet'' in astronomy papers, which is $g \sim 0.11 \pm 0.003$. This shows that slang, on average, grows significantly more rapidly than terminology within the field, indicating an added layer of difficulty for inclusion in future publications. To prepare for future publications, we fit for a date at which slang would comprise a majority ($> 50\%$) of an average Astronomy and Astrophysics publication.

\begin{figure}
    \centering
    \includegraphics[width=1.0\linewidth]{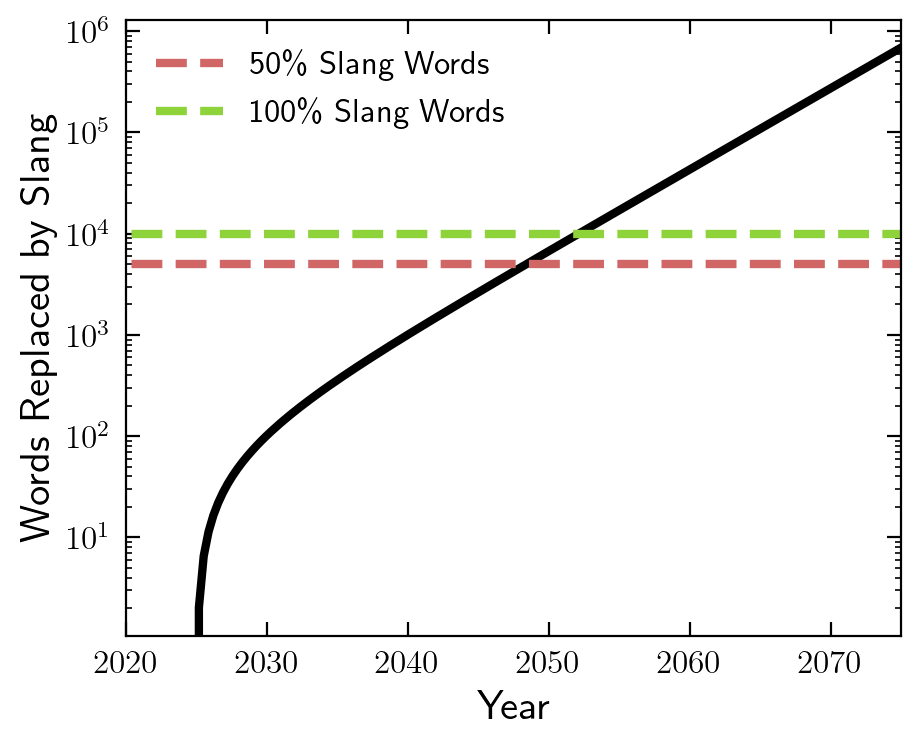}
    \caption{We model the replacement of traditional academic words with of slang words as an exponential with a growth rate of $0.18$ beginning in 2025 (solid black line). If we assume that an average Astronomy and Astrophysics paper has 10,000 words, by 2048 50\% of the words will be replaced by (red dashed line), and by 2052 100\% of the paper will be slang (green dashed line).}
    \label{fig:words_to_slang}
\end{figure}

First, we assume that the average Astronomy and Astrophysics publication is double-column in structure, and 10 pages long.
This allows for an average (in fact a high-end estimation as it excludes formatting of the title, abstract, and citations) of 10,000 words.
Our second assumption is that many of you are moved by this publication and begin to incorporate slang into your own.
This sets the zero-point for our function as 2025.
We now model the inclusion of a new slang words as our predetermined exponential with a growth rate $0.18$.
Figure \ref{fig:words_to_slang} shows the repercussions of replacing traditional academic words with slang words at this modeled rate. 
By April 2048, 50\% of the average Astronomy and Astrophysics publications will be comprised of slang words, and by 2052 100\% will be slang. 

\section{Applications}
\label{sec:apply}
It is a daunting task for any educator attempting to communicate with the youth of today in their ``own language". As such, we have assembled several example paragraphs for the reader to further introduce themselves with applications to the phrases in Tables \ref{table:Slang_Positive}, \ref{table:Slang_Neutral}, and \ref{table:Slang_Negatve}. Several passages from influential astronomy manuscripts have been translated into Gens $\alpha$ and Z terminology. By providing a direct comparison between these ``translated" versions and the originals, we hope to showcase the accessibility of this package and encourage further translation attempts throughout Academia. We stress that the terminology we provide is not only applicable to astronomical topics, but should be encouraged across all scientific fields to encourage readership and understanding.
\begin{itemize}
    
    \item \textbf{Gravitational wave detection:}
    
    ``In 1916, the year after the final formulation of the field equations of general relativity, Albert Einstein predicted the existence of gravitational waves. He found that the linearized weak-field equations had wave solutions: transverse waves of spatial strain that travel at the speed of light, generated by time variations of the mass quadrupole moment of the source [1,2]. Einstein understood that gravitational-wave amplitudes would be remarkably small; moreover, until the Chapel Hill conference in 1957 there was significant debate about the physical reality of gravitational waves." \citet{abbott2016}.

    In 1916, the year after King cooked the On-God Gyatt of the field equations of general relativity, Albert Einstein cooked the existence of gravitational waves. He High-Key'd that the skibidi weak-field equations had wave solutions: transverse waves of spatial opps that crash out at the speed of light, cooked by time squabbles of the mass quadrupole moment of the source. Einstein No-Capped that gravitational-wave amplitudes would be Low-Key; moreover, until the Chapel Hill conference in 1957 there was significant opps about the sus lore of gravitational waves. 

    \item \textbf{Hubble's Constant:}

    ``The major conclusion is that there is no reason to discard exploding world models on the evidence of inadequate time scale alone, because the possible values of H are within the necessary range." \cite{sandage1958}

    The High-Key conclusion is that, no-cap, we can't discard the exploding world models as we lack mad evidence of sus time scale, because of the On-God H values.

    \item \textbf{Tully-Fisher:}

    ``We propose that for spiral galaxies there is a good correlation between the global neutral hydrogen line profile width, a distance-independent observable, and the absolute magnitude (or diameter)." \cite{TullyFisher1977}

    We rizz you there is no cap for the the gucci relationship of spiral galaxies and the hydrogen line gyatt, not sus of distance, and absolute magnitude.

    \item \textbf{First exoplanet discovery:} 
    
    ``The results described here strongly suggest that one of the nearby galactic millisecond pulsars, PSR1257 + 12, is accompanied by a system of two or more planet-sized bodies." \cite{Wols1992}

    The results no-cap show the aura of the nearby galactic millisecond pulsars, PSR1257 + 12, is gucci, and the system may be cooking with two or more high-key planet-sized bodies - original GOAT planets.

    \item \textbf{Thin disk accretion:}
    
    ``The black hole (collapsar) does not radiate either electromagnetic or gravitational waves. Therefore, it can be found only due to its influence on the neighboring star ..." \cite{salpeter64, zeldovich64, zeldovich71, SS1973}

    The light opp has mad low-key radiation in either electromagnetic or gravitational aura. No cap, it can be made high-key only because of its Ohio rizz on the neighboring star...

\end{itemize}

\section{Conclusions}
\label{sec:conclsion}

While we recognize the importance of outreach and communication with younger generations, our results demonstrate that most children find it ``cringe" when adults or authority figures attempt to extend their vocabulary into the domain of modern slang. Our results indicate that, while young people can comfortably use slang to communicate with other young people, this rapidly loses efficacy if the speaker is aged 30 years or older (Priv. Comm., Author's nine-year-old niece). This perceived ``cringe", paired with an untenable growth of slang, casts doubt on the efficacy of utilizing slang as a tool for outreach. When we also consider that much of the slang discussed begins with marginalized groups and is later stripped of context (and let's be honest you, reading this, don't know where the slang comes from), we do not recommend artificially including such phrases in current publications, instead leaving the language to change naturally. Moreover, our predictions indicate that, at the current rate of slang development, paper posted to {\sc ar$\chi$iv} would be up to 50\% slang, and likely incomprehensible by a reader, after 2048.

We recommend further observations of the youth or those chronically online slowly devolving into total brainrot. 
\noindent
 
\noindent
No cap.

\bibliography{main}{}
\bibliographystyle{aasjournal}

\end{document}